# IBIC analysis of a linear position sensitive detector: model and experiment

Ettore Vittone (ORCID: 0000-0003-3133-3687)
*Physics Department, and and "NIS" Inter-departmental Centre, University of Torino, via P. Giuria 1, 10125 Torino, Italy*

Georgios Provatas (ORCID: 0000-0001-9471-1572)
Karla Ivanković Nizić (ORCID: 0009-0005-2177-8297)
Milko Jaksic (ORCID: 0000-0002-6750-9990),
*Division of Experimental Physics, Ruđer Bošković Institute, Bijenička cesta 54, 10000 Zagreb, Croatia*

Corresponding author: ettore.vittone@unito.it

## Abstract

The ion beam-induced charge (IBIC) analysis of a commercial silicon photodiode configured as a Position Sensitive Detector (PSD) for energetic charged particles is the subject of this report.

Although the photodiode is designed for detecting the position of incident light and optimized for use in the UV region, we present evidence that it also performs well as a detector for ions, with energies in the MeV range.

The device consists of a uniform p-type layer formed on a high-resistivity n-type semiconductor substrate, a pair of electrodes on both ends of the resistive layer, and a common electrode located on the backside of the substrate.

The IBIC experiment was carried out at the Laboratory for Ion Beam Interaction (LIBI) of the Ruder Boskovic Institute in Zagreb (HR), using 2 MeV proton microbeam raster scanning the 2.5x0.6 mm$^2$ active area of the PSD.

Each of the three electrodes was connected to independent standard NIM charge-sensitive electronic chain and the induced charge pulses associated with the position of individual ions were then digitized using a multi-channel analyzer interfaced with the SPECTOR software.

The longitudinal Charge Collection Efficiency (CCE) profiles acquired from the top electrodes show linear behaviors with opposite slopes, whereas the profile relevant to the signals from the back electrode is almost constant.

A model based on the IBIC theory satisfactorily interprets these results. It offers an alternative viewpoint to the commonly adopted Lateral Effect Photodiode principle and paves the way for the development of new PSDs for the identification of the impact position of a MeV ion with a resolution at the micrometer scale.

## 1. Introduction

Among the many types of devices capable of measuring the position of a light spot, position semiconductor sensors based on lateral photo effect [1] play an important role in various applications spanning from laser-tracking to synchrotron x-ray beam profiling [2].

A very common design of position sensitive devices (PSDs) consists on a photodiode with a resistive layer on the illuminated side with two (1D) or four contacts (2D) and a common electrode on the backside. The photocurrent gets split in the resistive layer in proportion to the resistance between the generation site and the contacts and the position of the incident light can be directly calculated from the ratio of the photocurrents.

The same devices have been investigated also for the measurement of position and energy of energetic charge particles [3–5]. In this case, the signal from the backside electrode is proportional to the particle energy, whereas the position signals are obtained from the charge fractions induced at the electrodes of the front-side resistive layer. These PSD's have found applications in Rutherford Backscattering Spectrometry/Channeling experiments [6] and for single electron/positron detection for emission channeling from radioactive atoms incorporated in crystals [7].

Ion Beam Induced Charge (IBIC) technique is an ideal technique to characterize such detectors [8]. Actually, charge signals from the sensing electrodes are detected synchronously with the incident position of ions, which are focused and scanned on the active area, in order to provide charge collection efficiency maps, which qualify the spatial sensitivity of the detector.

Noteworthy, IBIC technique is supported by a robust theoretical background [9], which models the relation between the incident particle position and the induced charge at the sensing electrode, in an alternative o complementary way of the commonly adopted charge division model [10,11].

In this work, an experiment was designed and conducted to characterize a commercially available position sensitive photodiode by the IBIC technique. The experimental results, satisfactorily interpreted by a theoretical model based on the IBIC theory, demonstrate that this type of detector is suitable to be used as a position sensitive detector for charged particles.

## 2. The experiment

The device under study is a SiTeK Position Sensing Detector 1L2.5 UV [12]. The device is made of a resistive p-doped layer on the front side fabricated on an n-type substrate (Figure 1). Two gold contacts are formed on the top surface, which was exposed to the proton µ-beam. The active area is (0.6x2.5) mm$^2$ and the resistance between these two electrodes is 50 kΩ.

A third electrode is on the back surface (ohmic contact), which was connected to a bias voltage power supply. The total capacitance of the device at 20 V of reverse voltage was less than 1 pF; the leakage current was around 10 nA.

The three electrodes were independently connected to charge sensitive preamplifiers and shaping amplifiers (shaping time = 1 µs) as shown in Figure 1. The three independent electronic chains converged to the Data Acquisition and Control system SPECTOR [13].

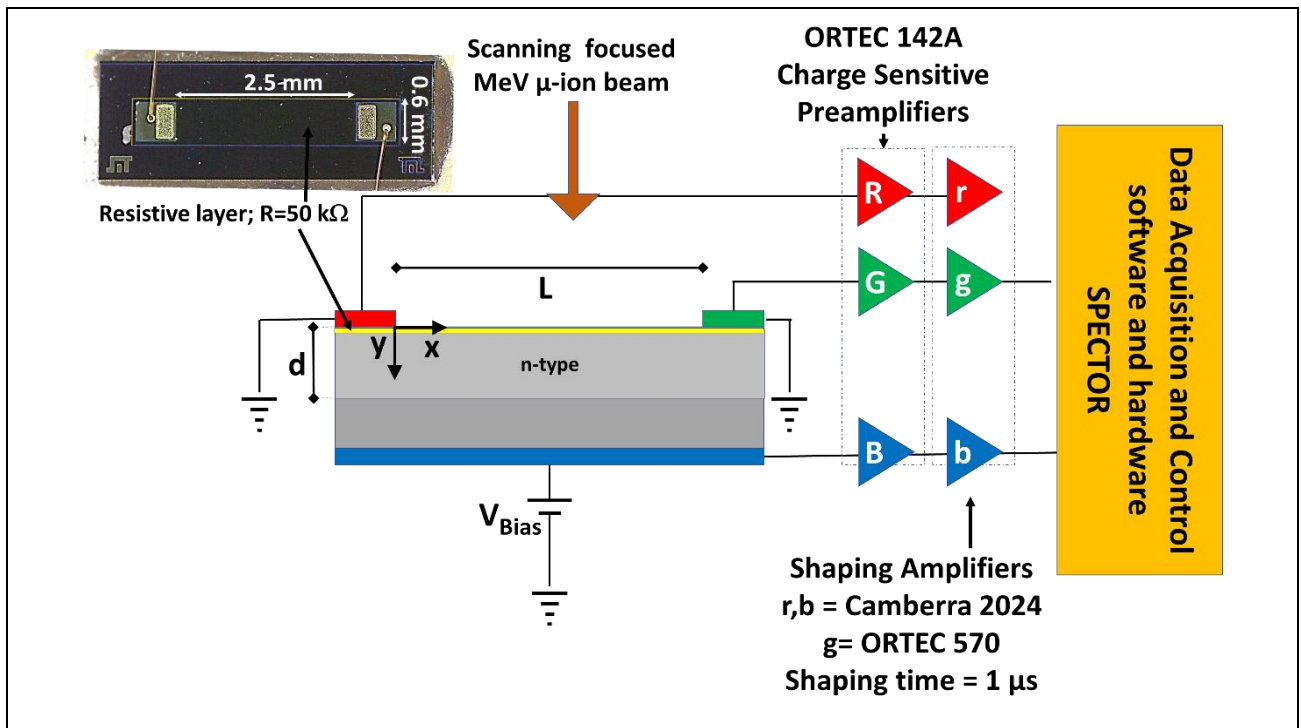

Figure 1: top left: image of the PSD. The scheme represents the cross section of the device and the electronic chain. The top surface was irradiated with the scanning focused proton beam.

IBIC experiment was carried out at the RBI (Ruđer Boskovic Institute) ion microprobe facility, using 2 MeV proton microbeam focused to a spot size of about 2 µm as estimated from by an on-axis scanning transmission ion microscopy (STIM) image of a copper grid (2000 mesh). The calibration of the electronic chain was performed by using a reference ORTEC Si PIPS detector, resulting in a charge sensitivity of about 3400 electron/channel [14].

For each of the three electrodes (the two top electrodes were grounded; the back electrode was biased at 20 V) IBIC maps were obtained by acquiring the relevant induced charge signals along with the coordinates of the proton microbeam, which raster scanned an area of (0.62x0.62) mm$^2$ of the device.

Eight different regions (128x128 pixels), were scanned to provide the IBIC map of the whole active area of the PSD, as shown in Figure 2. Each of them was obtained after shifting the sample by 500 µm from the previous position using the SmarAct xyz nanopositioning Piezo stage.

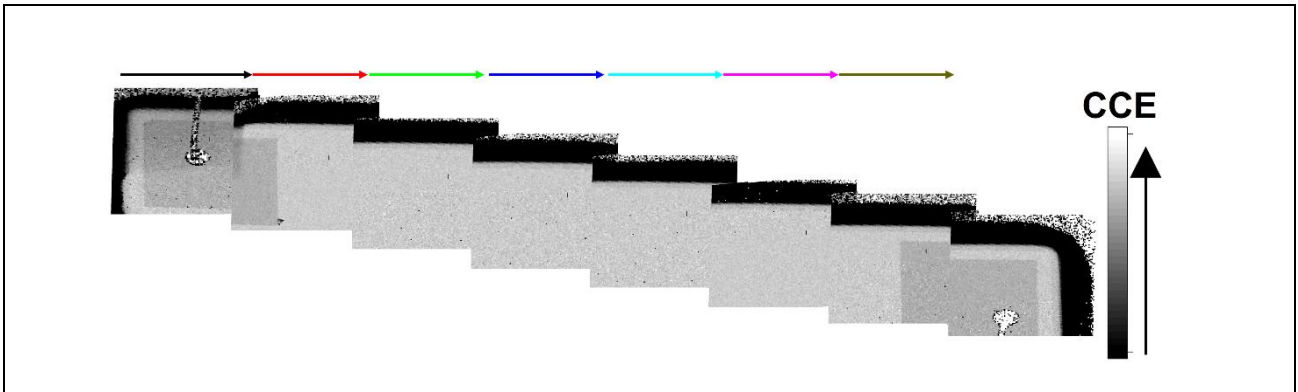

Figure 2: assembly of the eight scan areas to cover the whole active area of the PSD. The colored arrows on the top indicate the 500 µm horizontal shift of the sample with respect to the left border of the scan area. The vertical shift serves as a guide for the eye. The CCE relevant to the back electrode is encoded in the grey scale shown on the right side.

## 3. Experimental results

Figure 3 summarizes the main results of this experiment. Part a) refers to the CCE map measured considering the top left electrode (ref. Figure 1), easily localized by the shadow of the wire bonding on the left side. The graph at the bottom shows the profile extracted from the region highlighted by the two white lines in the map. The color of the markers identifies the eight scan regions. It is apparent that at the left electrode the CCE is maximum and linearly decreases as a function of the distance from such electrode, as shown by the oblique line, which represents the linear fit of the CCE in the active region. The linear regression provides a slope of $(-3.270\pm0.005)10^{-4}$ µm$^{-1}$.

The maps and profiles of Figure 3b are specular to those of Figure 3a, referring to the right top electrode. For this sensing electrode, the slope of the linear fit is $(3.438\pm0.004)10^{-4}$ µm$^{-1}$

Figure 3c shows the map and profile of the CCE measured from the back electrode. In this case the CCE profile is almost constant and, noteworthy, it is the sum of the profiles from the two top electrodes.

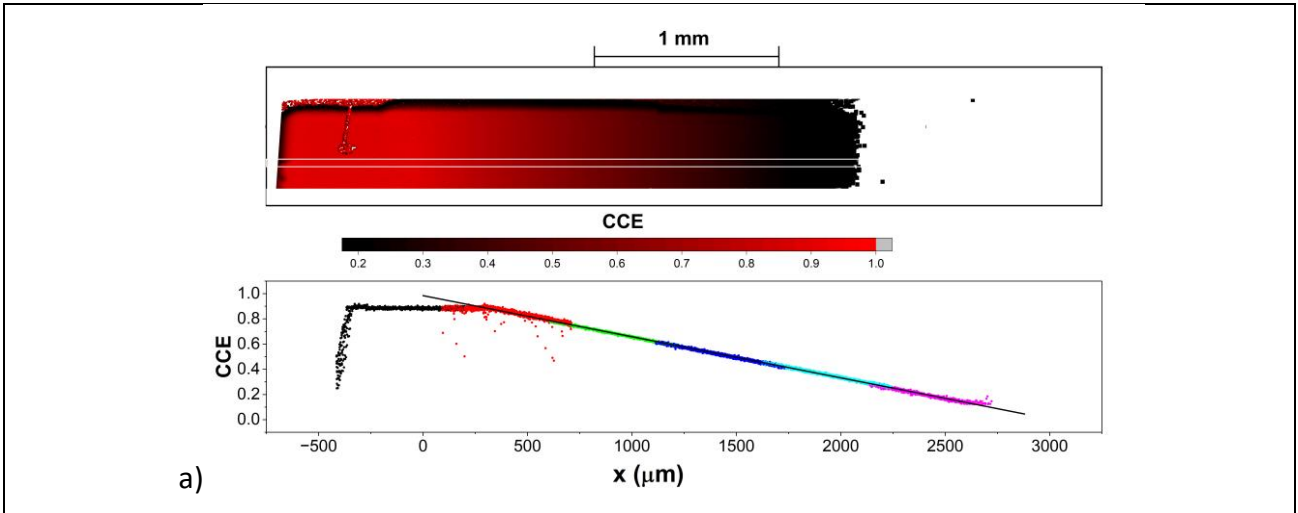

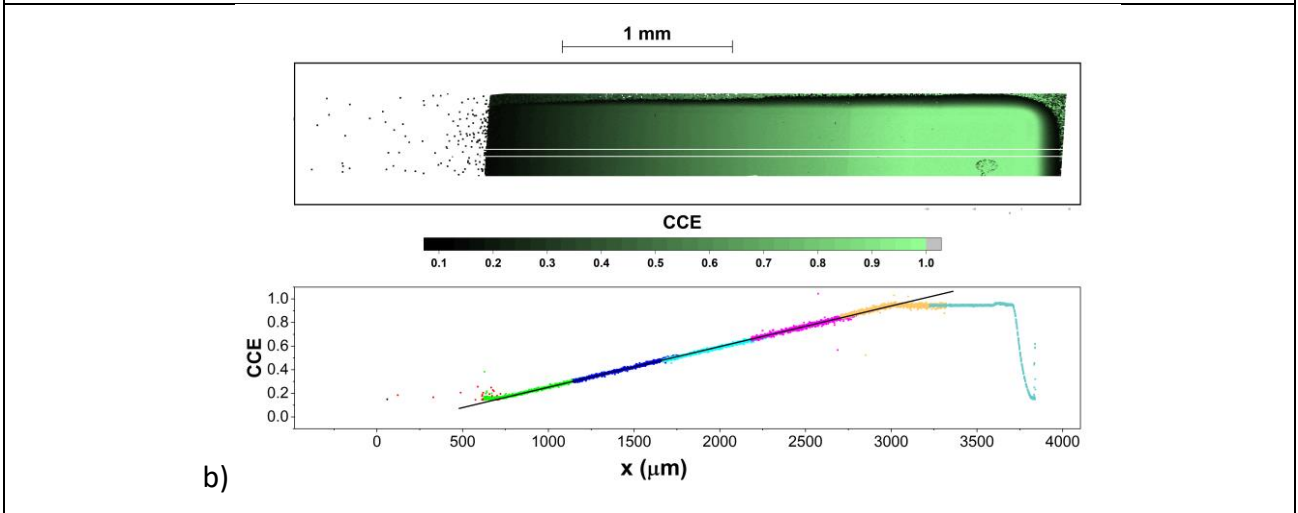

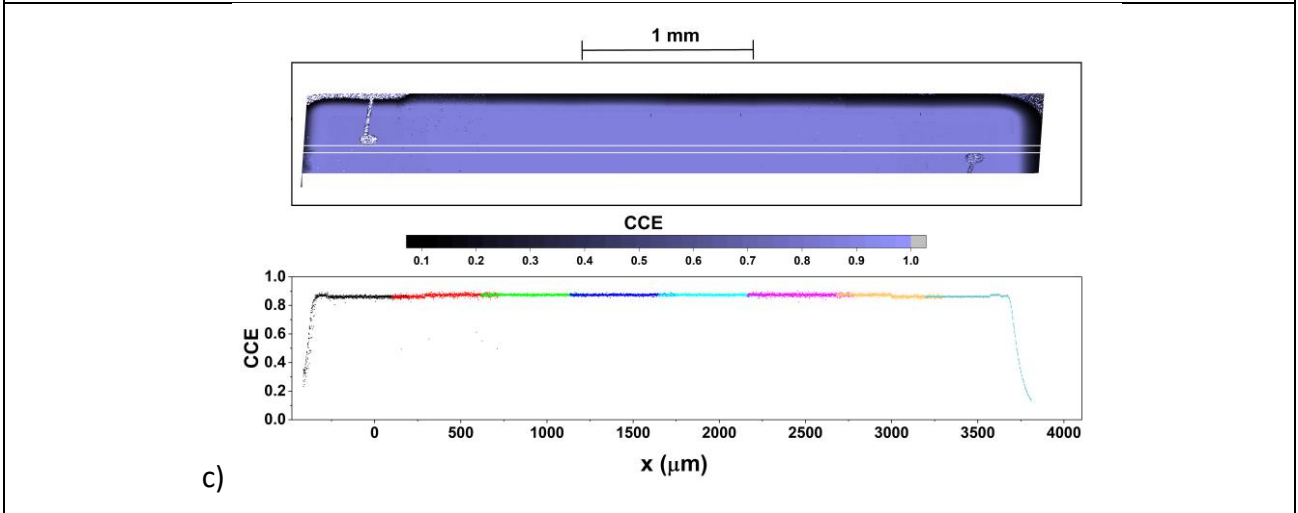

Figure 3: CCE maps (top) and profiles (bottom) referred to (a,b) the two top electrodes and c) to the back electrode (ref. Figure 1). The CCE is encoded in the horizontal scales. The white stripes on the maps indicates the region along with the profiles have been extracted. The colors of the markers in the profiles refer to the eight scan regions to cover the whole active area. The black line in figs a) and b) refer to the linear fit calculated in the central region, excluding the electrodes.

## 4. The model

The experimental results presented in the previous section can be satisfactorily interpreted by a model based on the IBIC theory [15], which is remarkably different from the theory of operation based on the charge dividing mechanism, which describes the PSD as a homogeneous line with a continuously distributed junction capacitance and resistance or on the diffusion time method [10].

In order to implement the model, we have to consider the electrostatics of the device to be evaluated in the rectangular domain of height d and length L (ref. Figure 1).

Being the capacitance at an applied bias voltage of 20 V smaller than 2 pF, the depletion layer width is larger than 100 µm, i.e. much larger than the 2 MeV proton range in Si (about 50 µm) and, for the purpose of this model, we can consider the extension of the depletion region equal to the depth of the domain d.

Assuming the space charge concentration ($\rho$), given by ionized donors in the n-type layer, constant along the x-axis in the depletion layer, the electrostatic potential $\Psi$ can be evaluated by solving the 1D Poisson's equation along the y coordinate:

(1) $\frac{d^2\Psi}{dy^2} = -\frac{\rho(y)}{\epsilon}$

where $\epsilon$ is the silicon dielectric constant, with the following boundary conditions:

(2) $\Psi(x, y = 0) = 0;\ \Psi(x, y = d) = V$

It is not within the scope of this paper to go deeper in the calculus of the electrostatic potential, which can be performed through standard numerical methods [16], nevertheless, it is worth stressing that $\Psi$ resulting from the solution of eq. (1) through the boundary conditions (2) is not a function of the x variable and the motion of charge carriers generated by ionization is strictly along the y axis, which is the direction of the electric field.

On the other hand, the Gunn's theorem [17–19] states that a point charge q moving from the initial position ($x_0,y_0$) to the final position ($x_1,y_1$) induces a charge $Q_i$ at each sensing electrode, i.e. i=R, G and B, given by

(3) $Q_i[(x_1, y_1), (x_0, y_0)] = -q[\Phi_i(x_1, y_1) - \Phi_i(x_0, y_0)]$

Where

(4) $\Phi_i = \frac{\partial \Psi}{\partial V_i}$

is the Gunn's weighting potential, which is defined as the derivative of the potential $\Psi$ with respect to the voltage at the sensing electrode $V_i$, assuming that all the other electrodes are grounded.

The weighting potential $\Phi_i$ is then the solution of the Laplace equation

(5) $\frac{d^2\Phi_i}{dy^2} + \frac{d^2\Phi_i}{dx^2} = 0$

resulting from the Poisson's equation (1), in which each term is derived by $V_i$; according to L-A Hamel et al. [20], we have assumed that the space charge does not contribute to the weighting field in the fully depletion conditions.

For the three sensing electrodes, the weighting potentials are then calculated by solving eq. (5) with the relevant boundary conditions given in Table 1

| Electrode | Boundary conditions | Solution of eq. (5) | CCE |
|---|---|---|---|
| R | $\Phi_R(x, y=0) = \left(1 - \frac{x}{L}\right)$ <br> $\Phi_R(x=0, y) = \left(1 - \frac{y}{d}\right)$ <br> $\Phi_R(x=L, y) = 0$ <br> $\Phi_R(x, y=d) = 0$ | $\Phi_R(x,y) = \left(1 - \frac{x}{L}\right) \cdot \left(1 - \frac{y}{d}\right)$ | $CCE_R(X) = \left(1 - \frac{X}{L}\right)$ |
| G | $\Phi_G(x, y=0) = \frac{x}{L}$ <br> $\Phi_G(x=0, y) = 0$ <br> $\Phi_G(x=L, y) = \left(1 - \frac{y}{d}\right)$ <br> $\Phi_G(x, y=d) = 0$ | $\Phi_G(x,y) = \frac{x}{L} \cdot \left(1 - \frac{y}{d}\right)$ | $CCE_G(X) = \frac{X}{L}$ |
| B | $\Phi_B(x, y=0) = 0$ <br> $\Phi_B(x, y=d) = 1$ <br> $\left.\frac{\partial \Phi_B}{\partial x}\right|_{x=0} = \left.\frac{\partial \Phi_B}{\partial x}\right|_{x=L} = 0$ | $\Phi_B(x,y) = \frac{y}{d}$ | $CCE_B = -1$ |

Table 1: Boundary conditions and solutions of eq. (5) and charge collection efficiency calculated through eq. (9) relevant to the electrodes R,G,B

Equations (3) relate to the induced charge generated by a single elementary charge moving inside the semiconductor. However, the electron-hole generation profile can be considered coincident with the ion stopping power versus depth, referred to as a Bragg curve and the carrier generation volume can be assumed to be as a function of the y coordinate only and nearly cylindrical with a diameter of tens of nanometers [8].

Therefore, for an ion incident on the sensitive area at the position x=X, the total induced charge at the i-th sensing electrode is then given by the superposition of the contributions of electrons and holes generated along the ion trajectory, that is, by the convolution of the generation profile Γ(y) and the charge induced by a single carrier moving inside the semiconductor, namely, for electrons:

(6) $Q_i^e(X) = +q \int_0^R dy \{\Gamma(y)\} \cdot [\Phi_i(X, 0) - \Phi_i(X, y)]$

And for holes:

(7) $Q_i^h(X) = -q \int_0^R dy \{\Gamma(y)\} \cdot [\Phi_i(X, d) - \Phi_i(X, y)]$

Where R is the ion range. The total charge induced at the sensing electrode is then given by:

(8) $Q_i(X) = Q_i^e(X) + Q_i^h(X) = +q \cdot [\Phi_i(X, 0) - \Phi_i(X, d)] \cdot \int_0^R dy\, \Gamma(y)$

And the relevant charge collection efficiency is given by the ratio of the induced charge Q and the total charge generated by ion incident at position X:

(9) $CCE_i(X) = \frac{+q \cdot [\Phi_i(X,0) - \Phi_i(X,d)] \cdot \int_0^R dy \Gamma(y)}{+q \cdot \int_0^R dy \Gamma(y)} = [\Phi_i(X, 0) - \Phi_i(X, d)]$

It is worth noticing that the sum of the CCE's is 0, i.e. the sum of the CCE's of the top grounded electrodes (namely R and G) is equal to $CCE_B$ (with the reversed sign). This conclusion is coherent with the hypothesis of full collection of the induced charge, which typically occurs in detector-grade materials, when the drift length of both carriers is much longer than the extension of the depletion layer.

For the two top electrodes, the weighting potential at y=d is null and, therefore, the relevant CCE's coincide with the boundary condition at x=0, namely $CCE_R = \Phi_R(X, 0) = 1 - \frac{X}{L}$ and $CCE_G = \Phi_G(X, 0) = \frac{X}{L}$,

On the other hand, the weighting potential at the top surface (y=0) is null, whereas at the back electrode is equal 1, and therefore $CCE_B = -\Phi_B(X, d) = -1$.

## 5. Discussion and Conclusions

The model outlined in the section 4. assumes full collection of the carriers generated by ionization by the ion probe, which is a reasonable assumption if the depletion layer width is larger than the ion range and the drift time (of the order of ns) is remarkably shorter than the carrier lifetime. These are commonly adopted assumptions for semiconductor detectors, where carrier lifetime typically exceeds microseconds [16].

The charge induced at each sensing electrode has been evaluated by applying the basic theory of IBIC, through the concept of weighting potential.

Since the weighting potential is a measure of the electrostatic coupling between the moving charge and the sensing electrode [21], the different shapes of the CCE's profiles relevant to the three electrodes are due to the coupling of the $\Phi_i$ and $\Psi$ maps.

The weighting potentials were calculated, by solving the Laplace's equation eq. (5), using the boundary conditions listed in Table 1,

The weighting potential map relevant to the back electrode $\Phi_B$ overlaps with the electrostatic potential $\Psi$ map; actually, assuming uniform doping distribution along the x-direction, the electric field and the weighted field are parallel to the y direction, resulting in a constant (along x) charge collection efficiency, in good agreement with the experiment (Figure 3c). The smaller value of CCE (about 90%) compared to the theory (100%) is to be attributed to the presence of the dead layer on the top surface mainly due to the resistive layer.

As regards the two top electrodes, the $\Phi_R$ and $\Phi_G$ maps are different from that of the electrostatic potential: actually, the solutions of eq. (5) are hyperbola for both the R and G electrodes, whereas the solution of eq. (1) is a function of only the y coordinate.

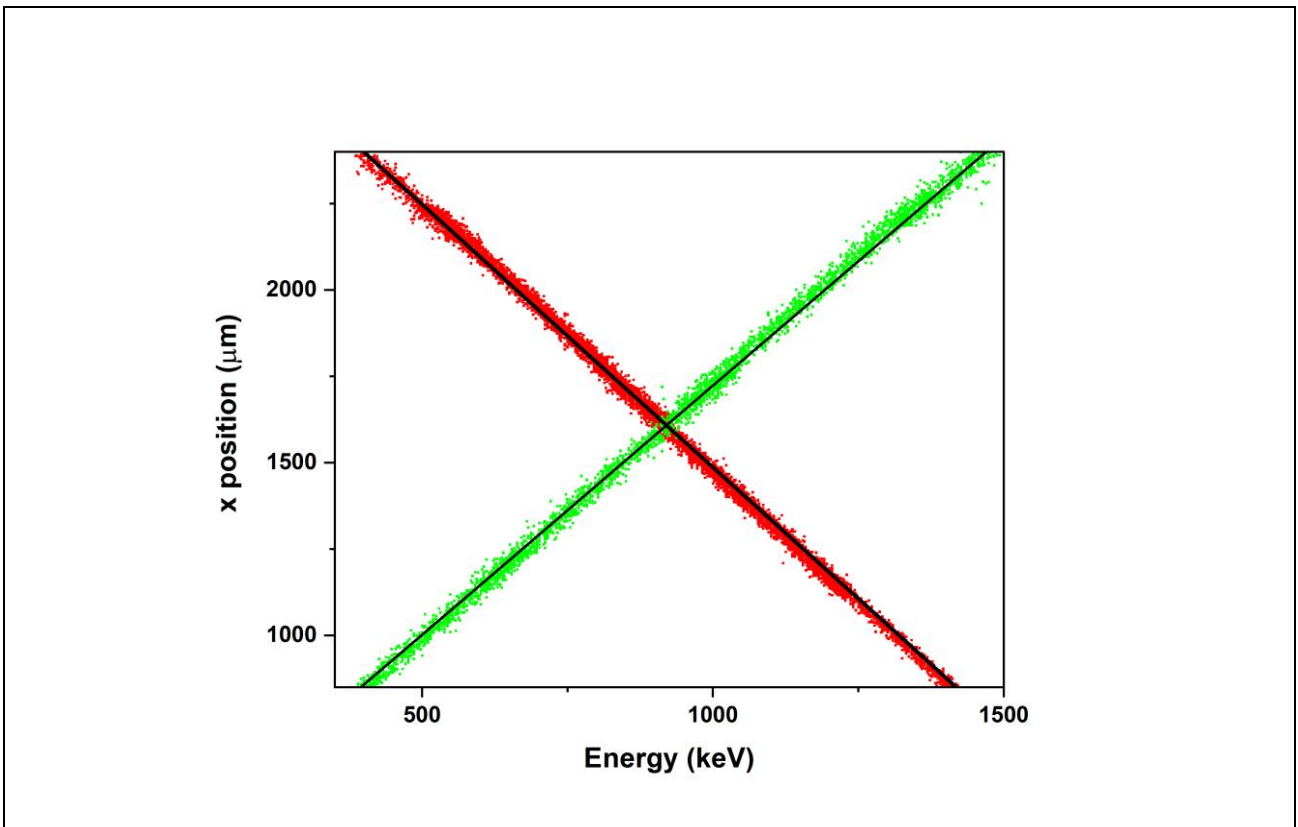

Figure 4: horizontal (x) position of the ion beam as a function of the signal measured by the two (R and G) top electrodes expressed in keV.

This is the main motivation of the specular and linear behaviors of the CCE's profiles shown in Figure 3a,b, which follow the linear potential drop between the R and G top electrodes.

The model satisfactorily interprets the experimental results described in section 3, which delineate the application of the device under study as a position sensitive charge particle detector.

In fact, the combination of the induced charge signal from in a single electrode provides the position of the ion beam (see Figure 4) with a resolution of 47 μm. The main source of the uncertainty in the evaluation of the beam position is given by the spectral resolution of our detection system, which is 32 keV from protons of energy 2000 keV, whereas the contribution of the fitting parameters is negligible. If the signals of both the electrodes are considered, the position resolution decreases to 34 μm.

In conclusion, our results demonstrate that this type of detector can be used as a position sensitive detector for 2000 keV proton beams, showing a good position linearity and a resolution of about 1.5% of the active area longitudinal axis. The model developed to interpret the experimental results corroborates the possibility that such devices, originally designed for optical applications, can be used to monitor the features of ion beams with other ion species and energies, provided that the ion range falls within the depletion layer width, in order to guarantee full depletion conditions, for example in the nuclear physics experiments which require a large angular coverage and resolution, as well to monitor on-line the shape of ion beams.

## Acknowledgements

This work has been supported by the European Union's Horizon Europe Research and Innovation program under Grant Agreement No 101057511 (EURO-LABS).

## Data availability

The data that support the findings of this study are available from the corresponding author upon reasonable request. The authors have no conflicts to disclose.